\documentclass[pra,aps,twocolumn,showpacs,superscriptaddress]{revtex4-1}

\usepackage{epsfig}
\usepackage{mathrsfs}
\usepackage{amsmath}
\usepackage{float}
\usepackage{color}
\usepackage{epstopdf}

\begin{document}

\title{}

\maketitle

\newcommand{\ds}{\displaystyle}
\newcommand{\dd}{\partial}
\newcommand{\be}{\begin{equation}}
\newcommand{\ee}{\end{equation}}
\newcommand{\beq}{\begin{eqnarray}}
\newcommand{\eeq}{\end{eqnarray}}
\newcommand{\dt}{\ds\frac{\dd}{\dd t}}
\newcommand{\dz}{\ds\frac{\dd}{\dd z}}
\newcommand{\D}{\ds\left(\frac{\dd}{\dd t} + c \frac{\dd}{\dd z}\right)}

\newcommand{\w}{\omega}
\newcommand{\W}{\Omega}
\newcommand{\g}{\gamma}
\newcommand{\G}{\Gamma}
\newcommand{\E}{\hat E}
\newcommand{\s}{\sigma}
\newcommand{\bra}{\langle}
\newcommand{\ket}{\rangle}


{\bf Salih, Li, Al-Amri and Zubairy reply:}
In his comment \cite{Vaidman}  on our letter \cite{Salih}, Vaidman makes the following points: (1) He agrees that when the blockade is there and interference is destroyed in the inner Mach-Zehnder interferometers, fully counterfactual information is obtained; (2) He agrees that even when there is no blockade, ``the branch of the wave function of the photon reaching detector $D_1$ does not pass through the communication channel''; (3) He, however, argues that our protocol is ``not counterfactual for the values of the information bit corresponding to the absence of the blockade'' as, according to him, ``given a click at $D_1$, the probability for finding the photon by a nondemolition measurement of the projection operator on the transmission channel is one''. In the following we show that Vaidman's claim that the photon exists in the channel---which hinges on his own interpretation of quantum mechanics---is wrong.

Vaidman argues that it is a mistake to say ``the probability of finding a signal photon in the transmission channel is virtually zero.'' One might wonder: how did he arrive at a unit probability for finding the photon in the channel by carrying out a strong nondemolition measurement---while the maximum probability amplitude for the photon state $\left| \text{001} \right\rangle$, corresponding to the photon being in the channel, is $\sqrt{T_M}$, where $T_M$ is the almost-zero transmissivity of beam-splitters $BS_M$? His nondemolition measurement is in fact a series of measurements on all cycles---making it, as he notes, equivalent to Bob blocking the channel---but instead selecting the rare event (near zero probability) of detecting the photon in the channel, i.e. the protocol failing. Then again the rare event (near zero probability) that the photon ends up at $D_1$, rather than one of the $D_3$'s or $D_2$, is selected to give a probability of one for finding the photon in the channel given a click at $D_1$: an imaginative use of post-selection to turn a near zero probability into exactly one!

Next we move to his weak measurement argument. In order to discuss only the case when there is no blockade, we have simplified our original setup in Fig 1. Here the smaller interferometer has 50-50 beam splitters  $BS_N$ while beamsplitters $BS_M$ have reflectivity $R_M$. This contains the essential features of our protocol. Here path C corresponds to the transmission channel. 

The essence of our protocol is that we can choose the transformation properties of $BS_N$'s such that any photon sent into the smaller interferometer will cause  $D_3$ to click with unit probability.  This means that the probability of the photon existing at location E is zero. In such a situation, the outcome of the experiment will be completely independent of whether the path E is open or blocked. Thus a click at $D_1$ implies that the photon should have followed path A, and the probability of its existence in the public channel is zero. This leads to the counterfactual  behaviour discussed in our paper. We should emphasize that this result is a direct consequence of standard quantum mechanics and to contradict it, as Vaidman does, goes beyond standard quantum mechanics.

Vaidman's argument hinges on the fact that if we measure the weak value of the photon number at C, it is non-vanishing. He then concludes that the photon should be in the transmission channel with unit probability. The weak values at D and E, when detector $D_1$ clicks are however zero. Thus we have the  paradoxical situation: no photon enters the inner interferometer (within the circle) and no photon leaves it, but the photon exists in the transmission path C. This implies that -1 photon should exist on path B.

In \cite{Li}, we have explained this result in detail concluding that a quantum measurement of the weak value of the projection operator $|001\rangle\langle 001|$ (photon number at C) for the post-selected state $|100\rangle$ (click at $D_1$) disturbs the interference in the inner interferometer no matter how weak the interaction. (Reply in \cite{Reply}.) This leads to a non-zero amplitude at E. The probability amplitude at $D_1$ then results from an interference of two amplitudes, one corresponding to a passage through path A and another through path DCE. Thus Vaidman's claim concerning the measurement at C is not valid.

\begin{figure}[H]
\begin{center}
\includegraphics[height=4.50cm,width=0.45\columnwidth]{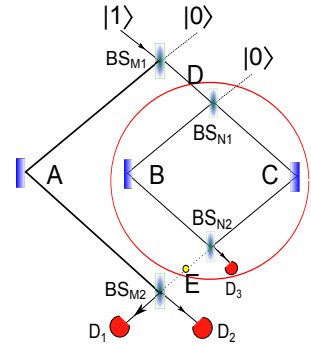}
\end{center}

\caption[2Dpattern]
   { \label{fig:fig1}
    The case of absence of a blockade.  }
\end{figure}

The mistake Vaidman makes is his implicit assumption that any weak measurement in arm C does not affect the interference in the inner interferometer, in direct conflict with the predictions of quantum mechanics. Our results in \cite{Li} prove this point through detailed analysis.

In summary, our analysis and claims in \cite{Salih} are all in accordance with the principles of quantum mechanics and we do not find any inconsistency in our conclusions.

\end{document}